\documentclass[journal,12pt]{IEEEtran}

\usepackage{comment}
\usepackage{soul}

\usepackage[dvipsnames]{color}
\usepackage[pdftex]{graphicx}
\graphicspath{{figures/}{Figures/}}

\usepackage{tikz}
\usetikzlibrary{calc}

\usepackage[cmex10]{amsmath}
\usepackage{amsfonts}
\usepackage{amsthm}
\usepackage{physics}

\usepackage{cite}
\usepackage{url}
\usepackage{tabularx}
\usepackage{booktabs}
\usepackage{multirow}
\usepackage[pdfpagelabels,hypertexnames=false,breaklinks=true,bookmarksopen=true,bookmarksopenlevel=2]{hyperref} 
\usepackage[capitalise]{cleveref}

\title{The Computational and Latency Advantage\\of Quantum Communication Networks}

\author{Roberto~Ferrara, 
        Riccardo~Bassoli, 
        Christian~Deppe, 
        Frank~H.P.~Fitzek 
        and Holger~Boche 
\thanks{R. Bassoli and F. H.P. Fitzek are with the Deutsche Telekom Chair of Communication Networks, Institute of Communication Technology, Faculty of Electrical and Computer Engineering, Technische Universität Dresden, Dresden, Germany. F. H.P. Fitzek is also with the Centre for Tactile Internet (CeTi) with Human-in-the-Loop, Cluster of Excellence, Dresden, Germany.
(e-mail: \{riccardo.bassoli, frank.fitzek\}@tu-dresden.de).}
\thanks{R. Ferrara, C. Deppe and H. Boche are with the Department of Electrical and Computer Engineering at Technische Universität München, München, Germany. H. Boche is also with the Munich Center for Quantum Science and Technology (MCQST), München, Germany.
(e-mail: \{roberto.ferrara, christian.deppe, boche\}@tum.de).}%
\thanks{{© 2021 IEEE. Personal use of this material is permitted. Permission from IEEE must be obtained for all other uses, in any current or future media, including reprinting/republishing this material for advertising or promotional purposes, creating new collective works, for resale or redistribution to servers or lists, or reuse of any copyrighted component of this work in other works.}}
}

\begin{document}
\markboth{Accepted for publication in the IEEE COMMUNICATIONS MAGAZINE}{Left}

\maketitle

\begin{abstract}
This article summarises the current status of classical communication networks and identifies some critical open research challenges that can only be solved by leveraging quantum technologies. By now, the main goal of quantum communication networks has been security. However, quantum networks can do more than just exchange secure keys or serve the needs of quantum computers. In fact, the scientific community is still investigating on the possible use cases/benefits that quantum communication networks can bring. Thus, this article aims at pointing out and clearly describing how quantum communication networks can enhance in-network distributed computing and reduce the overall end-to-end latency, beyond the intrinsic limits of classical technologies. Furthermore, we also explain how entanglement can reduce the communication complexity (overhead) that future classical virtualised networks will experience.
\end{abstract}

\section{Introduction}

    \IEEEPARstart{T}{he} history of telecommunications has already experienced a fundamental progress-driven paradigm shift from \textit{circuit switching} to \textit{packet switching}. However, the necessity to interconnect very heterogeneous networks and to target different verticals (mobile broadband, augmented reality, vehicular networks, Tactile Internet, Industry 4.0, etc.), with different concurrent and stringent requirements, have raised the need for a new generation of networks, called 5G and beyond~(B5G). The scope of these networks is to provide an ecosystem of networks flexibly, efficiently and effectively interconnecting heterogeneous radio access networks~(RANs) and wired networks (edge, core and the Internet). At the same time, requirements of very low latency, significantly greater throughput, increase of energy efficiency, and of ubiquitous connectivity have also pushed research and industrial community to investigate new paradigms for telecommunications.

	\begin{figure}
	    \centering
		\includegraphics[width=\columnwidth, trim = 0 0 0 60, clip]{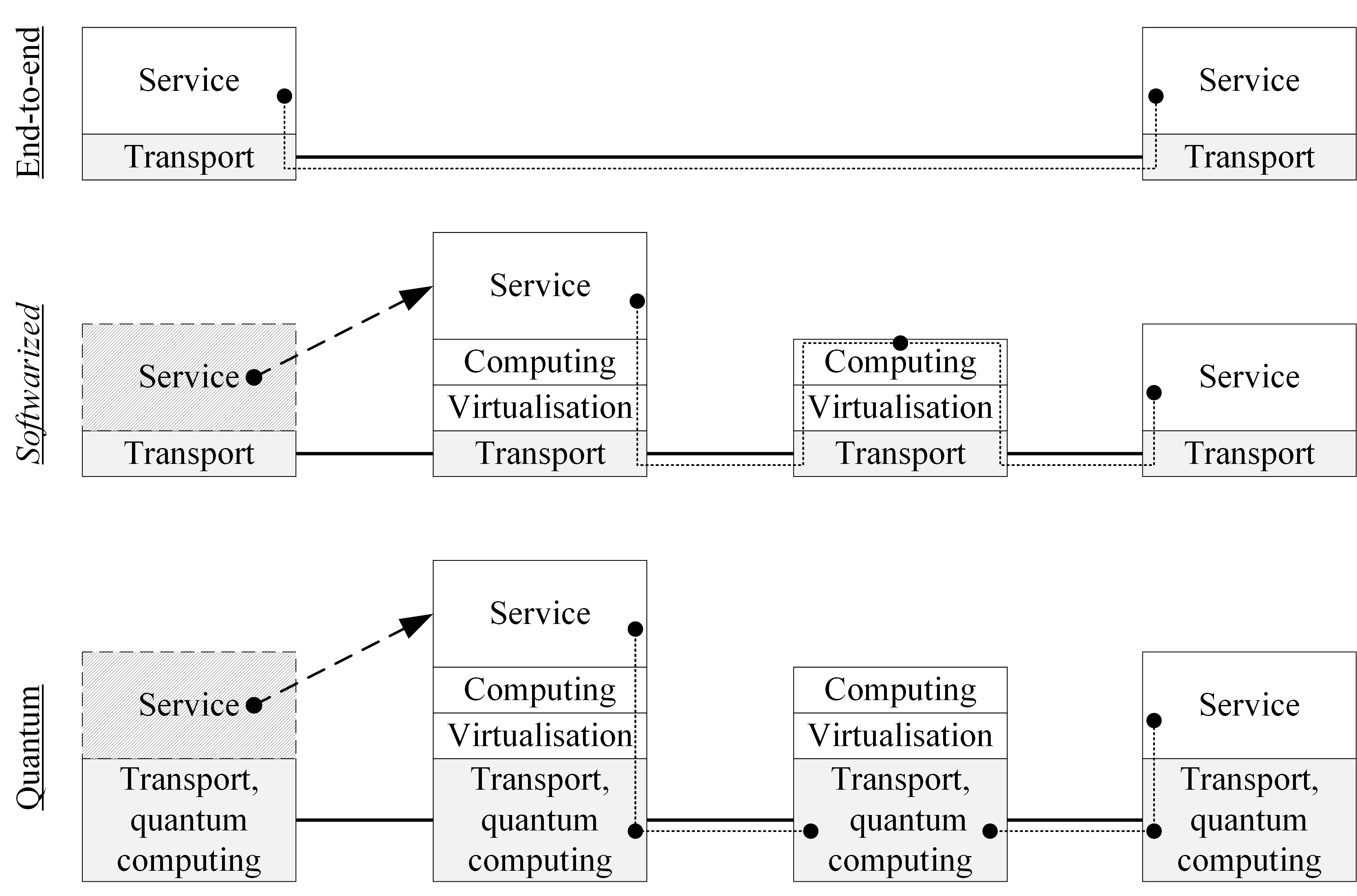}
		\caption{The paradigms of softwarized and quantum communication networks.}
	    \label{fig:paradigms}
	\end{figure}
	
	In order to solve these problems, since 2010 the idea of network \textit{softwarization} has become popular via software-defined networking~(SDN) and network function virtualization~(NFV), as a realization of the paradigm shift from \textit{store-and-forward} to \textit{compute-and-forward} (\cref{fig:paradigms}). 
	The idea behind such technologies is the transformation of communication networks from employing dedicated hardware to using general purpose hardware, running software-based network components (instantiated inside virtual machines, containers, unikernels, etc.). 
	Nevertheless, software-based future generation networks will not be able to satisfy the expectations, because of classical software and network-theoretic intrinsic limits, while they will also introduce some critical drawbacks. Firstly, the communication complexity (i.e. the number of bits exchanged among distributed network nodes to compute a function) of these networks, relying on massive distributed computing, will significantly increase.
	Secondly, the security of the software functions will be weaker than that of dedicated hardware, requiring a continuous growth of security level at all layers of the software-based network stack, which will add overhead, delay and resource usage. 

	A radical reform of the network nature is necessary to target existing and future expectations. This intrinsic reform can come from exploiting quantum-mechanical resources such as superposition of quantum states, quantum entanglement, and distributed quantum computing, beyond targeted quantum security use cases. Quantum networks will be built on top of classical ones, in a unique hybrid infrastructure, where \textit{quantum virtual machines} will consist of a high number of entangled spatially-distributed qubits and scaling with the number of interconnected devices. This hybrid classical-quantum communication network is normally called by the research community, the Quantum Internet \cite{CacciapuotiCaleffiTafuriCataliottiGherardiniBianchi2020}. 
	Generally, the goal of the Quantum Internet is enabling the transmission of qubits between distant quantum devices to achieve the tasks that are impossible using classical communication. However, the definition of the classical 5G and B5G networks is much broader.
	Since we do not want to develop quantum versions of the classical Internet components, the term Quantum Internet is rather misleading.
	In this work, we thus prefer to use the term \emph{quantum communication network}~(QCN)~\cite{QuantumBook2021}.
	
    Many quantum network projects have the only goal of securely exchanging cryptographic keys. Entanglement-based protocols like Ekert '91, as opposed to Bennet-Brassard '84 like protocols, are often used  because they work directly with quantum repeaters. Other popular quantum network projects focus on using the entanglement for teleportation. However, using the quantum network for just these applications is limiting its potential.
        
    \begin{figure*}
	    \centering
		\includegraphics[width=1.9\columnwidth]{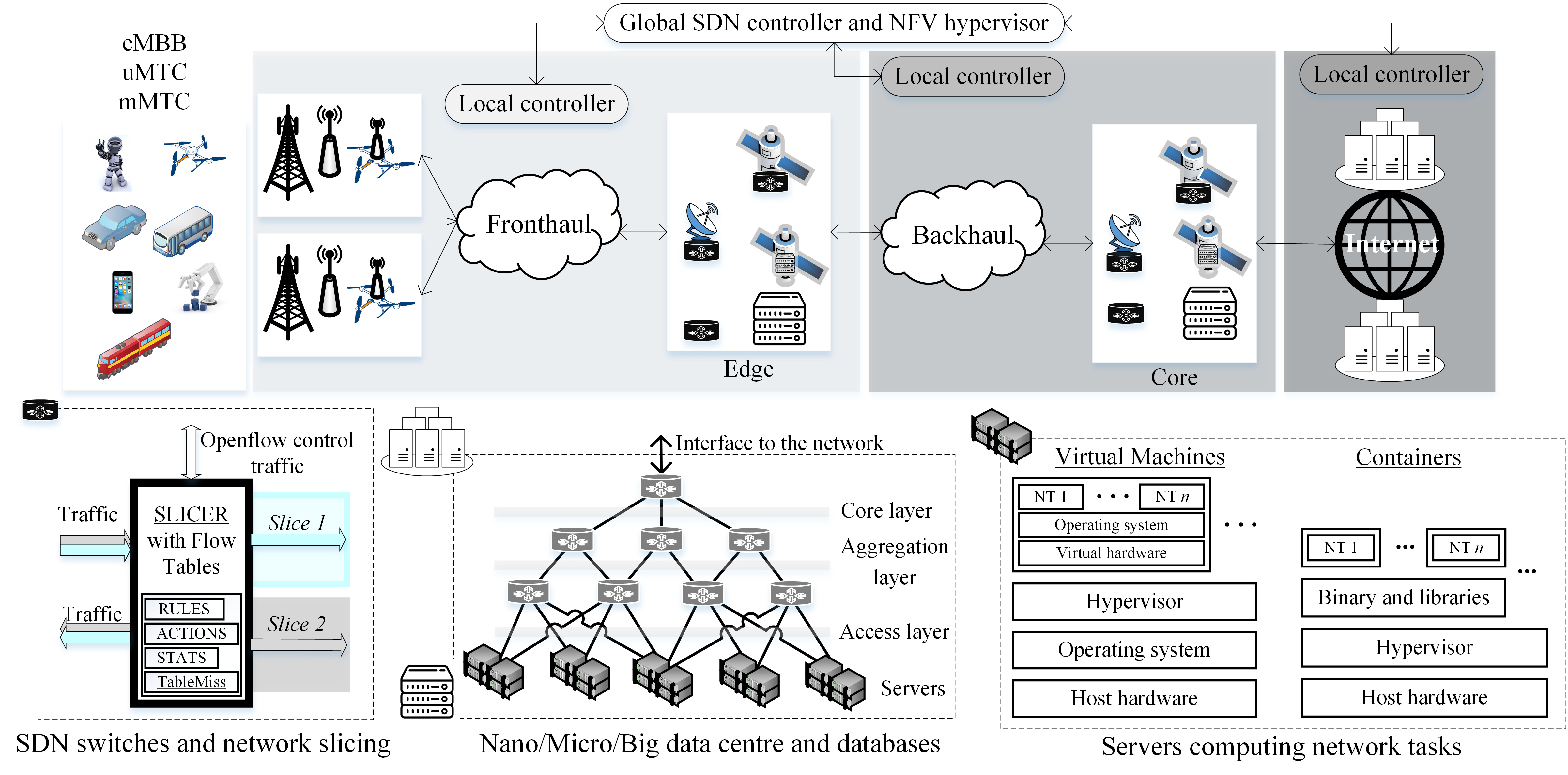}
		\caption{Logic architectural representation of classical future softwarized networks, with their main components. Local/global controllers update the routing tables and routing policies (RULES, ACTIONS and TableMiss) at the SDN switches via a control protocol (e.g. Openflow). SDN switches collect statistics (STATS) to improve network analysis at the controllers. The hypervisor will very likely collect and process Big Data to run machine learning algorithms for the whole network management.}
	    \label{fig:ClassicalNetworks}
	\end{figure*}

\section{In-Network Computing and Big Data} \label{sec:classicalcomputing}
    As previously mentioned, a critical aspect is \textit{low latency} \cite{JiangShokriGhadikolaeiFodorModianoPangZorziFischione2019}. Latency can be seen as the sum of different parameters, referred to all the network levels.
    The \textit{propagation latency} depends on spatial distance and packet size.
    While source data is constant, each layer of the network stack adds its own overhead, contributing to this delay.
    The \textit{transmission delay} at each link is proportional to the inverse of the available capacity.
    Next, the \textit{queuing delay} is due to the queues of packets at all the nodes of the network, which varies for different frames/datagrams/packets depending on their Quality-of-Service~(QoS).
    The \textit{processing delay} is the physical and link layer delay, which mainly depends on the hardware's processing capacity of network nodes and on the signal-processing algorithms.
    Currently, reduction of latency is done per layer.
    For example, at the physical layer propagation delay is reduced via the transmission of short packets at the cost of increasing  the overhead per frame and thus the transmission delay.
    However, when a single layer is overloaded, optimising the network per layer prevents the other layers to overtake some of the load efficiently, and may result in the whole network being overloaded.
    For example, the medium-access layer is responsible for synchronisation, initial access, interference management, scheduling, rate adaptation, delay-to-access and management of retransmission, 
    while the transport layer is responsible for the flow and congestion controls. 
    If congestion causes significant losses at the transport layer, it can increase queuing, processing and retransmission delays (losses increase both retransmission time and redundancy for error correction) at the medium-access layer.

    Future softwarized NFV-SDN communication networks will consist of a data plane and a control plane (\cref{fig:ClassicalNetworks}), which will be obtained by a resource mapping from  general-purpose hardware owned by various infrastructure providers (also called tenants). Network nodes and traffic in the data plane will be managed and processed by SDN controllers and hypervisor in the control-plane network, with the help of machine learning~(ML) towards self-organising (autonomic) networks~(SONs)~\cite{XieYuHuangXieLiuWangLiu2019}.
    Moreover, in the protocol stack, the number of active layers will be optimised at each network node improving latency.
    \cref{fig:ClassicalNetworks} depicts the logical architecture of future \textit{softwarized} classical networks, highlighting the internal structure of their main components. 
    The complete \emph{softwarization} of networks and the realisation of in-network intelligence will transform the network stack and structure from static to adaptive, through the constant processing of Big Data collected from the enormous amount of devices of the network.
    
    Sometimes these virtual-network optimisation problems can be efficiently distributed, allowing to distribute the overhead load and delays within the network by having distributed SDN controllers into various network nodes or as virtual network functions (VNFs) in different servers or data centres~\cite{BannourSouihiMellouk2018}.
    However, the flexibility introduced by \emph{softwarization} will inevitably come with an overhead trade-off, which will reduce the improvement in terms of latency and energy efficiency~\cite{CaoLiuChengShen2018}.
    This overhead can be potentially resolved, or greatly reduced, in a classical-quantum hybrid network, by exploiting the fact that even some classical tasks can be solved much more efficiently in the quantum network, rather than the classical one. All the previously mentioned delays are not only in the network but also in data centres and distributed databases, which are networks themselves (\cref{fig:ClassicalNetworks}).

\section{From Classical to Quantum} \label{sec:qcn}
    
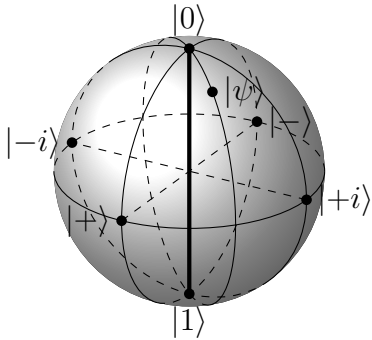
\begin{figure}
        \centering
        
        \newcommand\pgfmathsinandcos[3]{%
          \pgfmathsetmacro#1{sin(#3)}%
          \pgfmathsetmacro#2{cos(#3)}%
        }
        
        \def\Elevation{25} 
        \pgfmathsinandcos\sinEl\cosEl{\Elevation} 
        
        \newcommand\LongitudePlane[2]{%
            \tikzset{#1/.estyle={scale=\R, cm={cos(#2),sin(#2)*\sinEl,0,\cosEl,(0,0)}}
            }
        }
        \newcommand\LatitudePlane[2]{%
            \tikzset{#1/.estyle={scale=\R, cm={cos(#2),0,0,cos(#2)*\sinEl,(0,sin(#2)*\cosEl)}}
            } %
        }
        \newcommand\DrawLongitudeCircle[1]{
          \LongitudePlane{plane}{#1}
          \pgfmathsetmacro\angVis{atan(sin(#1)*\cosEl/\sinEl)} %
          \draw[plane,thin,black]  (\angVis:1) arc (\angVis:\angVis+180:1);
          \draw[plane,thin,dashed] (\angVis:1) arc (\angVis:\angVis-180:1);
        }%

        \newcommand\DrawLatitudeCircle[1]{
          \LatitudePlane{plane}{#1}
          \pgfmathsetmacro\angVis{asin(min(1,max(tan(#1)*\sinEl/\cosEl,-1)))}
          \draw[plane,thin,black]  (\angVis:1) arc (\angVis:-180-\angVis:1);
          \draw[plane,thin,dashed] (\angVis:1) arc (\angVis: 180-\angVis:1);
        }%
        

        \begin{tikzpicture}[
            scale=.45,
            dot/.style={inner sep=0pt,outer sep=0pt,minimum size=4pt, fill=black, circle},
            ]
            \def\R{4} 
        
            \fill[scale=\R, ball color=white!10] (0,0) circle (1); 
        
            \LongitudePlane{XZbasis}{-120}
            \DrawLongitudeCircle{-120}
            \draw[XZbasis, dashed] 
                (0: 1) coordinate[dot] (plus)  node[left]  {$\ket{+}$}
                --
                (0:-1) coordinate[dot] (minus) node[right] {$\ket{-}$}
            ;
        
            \LongitudePlane{YZbasis}{-30}
            \DrawLongitudeCircle{-30}
            \draw[YZbasis, dashed] 
                (0: 1) coordinate[dot] (plusi)  node[right] {$\ket{+i}$}
                --
                (0:-1) coordinate[dot] (minusi) node[left]  {$\ket{-i}$}
            ;
        
            \DrawLatitudeCircle{0} 
            \draw[XZbasis, ultra thick] 
                (90: 1) coordinate[dot] (zer) node[black, above]  {$\ket{0}$}
                --
                (90:-1) coordinate[dot] (one) node[black, below] {$\ket{1}$}
            ;
            
            \LongitudePlane{psiPlane}{-70}
            \DrawLongitudeCircle{-70}
            \draw[psiPlane]
                (60:1) coordinate[dot] (psi) node[right] {$\ket\psi$}
            ;
        \end{tikzpicture}

        \caption{The Bloch sphere of a qubit. It is the same as the Poincar\'e sphere of light polarisation but with the behaviour of a single photon. The orthogonal polarisation vector states $\ket{0}$ and $\ket{1}$ can be used to store a single classical bit. 
        Each point in the solid vertical line represents a probability distribution over one bit, with the centre being the uniform coin toss. New pure states are the boundary of the sphere, that is the analogue of deterministic classical states (the dots are some examples). 
        The inside of the sphere represents mixed states, probabilistic mixtures of antipodal pure states from the boundary.}
        \label{fig:qubit}
    \end{figure}

    Classically, the smallest information element is the bit, an object that stores information by using two possible values: $0$ and $1$. 
    In quantum systems bits do not exist.
    The state of any quantum-mechanical system shows a linear behaviour similar to that of continuous waves and signals\footnote{Thus the name wavefunction in quantum mechanics.}.
    The unit of quantum information is thus the \emph{qubit}, an object that stores information on the unit vectors of a two-dimensional complex vector space.
    The connection with the classical bit is that $0$ and $1$ now index the two standard basis vectors as $\ket{0}$ and $\ket{1}$\footnote{$\ket\psi$ is the Dirac's bra-ket notation for vectors.}. 
    Just like for continuous signals, the sum of $\ket{0}$ and $\ket{1}$ states is another state that is different from probabilistically generating either of the two. The standard representation of the state space of a qubit is displayed in \cref{fig:qubit}.

    The difference between a signal and a quantum state lies in their behaviour.
    The state of a qubit should be thought as the polarisation of a single photon, where the only collectable information is whether the photon is present or not. 
    We can distinguish two orthogonal states of the photon such as the vertical and horizontal polarisation using a polarising beam splitter, and detecting on which side the photon appears. A qubit can thus carry one bit of information.
    However, if we try to collect information about a diagonally polarised photon (with the same vertical/horizontal measurement), we will still register the photon only in either the vertical or the horizontal measurement.
    Only repeating the experiment with many photons, we probabilistically observe either polarisation, but never both simultaneously.
    Still, while in the above measurement the photon behaves probabilistically, there exists the diagonal-polarisation measurement in which the photon behaves deterministically.
    This probabilistic/deterministic behaviour of the qubit depends on how we measure it, and thus we refer to those states that behave deterministically in some basis as \emph{pure state}.
    The fact that measuring a photon only gives a single bit of information\footnote{This is not in contradiction with superdense coding, where one qubit carries two bits of information, but two qubits must be measured to obtain them.} implies that we cannot distinguish arbitrary pure states and, after the measurement, the original pure state is destroyed: the quantum information of arbitrary pure states cannot be cloned/copied.
    
    The different behaviour of qubits becomes even more drastic when we consider multiple qubits. There exist states that behave probabilistically under any measurement performed independently on each qubit, but deterministically under a joint measurement that can be performed only with global access to the involved qubits. These are the \emph{entangled pure states}.
    Now, it is important to clarify that the randomness produced by a pure state is fundamentally different from the randomness produced by a classical system: the classical system \emph{knows} the value of the randomness (which is simply hidden), while in a pure state not even the quantum system itself knows the value before the measurement is performed.
    Said otherwise, the randomness does not even exist before measuring the qubit.
    This is, for example, how Quantum Key Distribution~(QKD) protocols share keys that are claimed provable-secure against any adversary: the protocols prove that the keys at the receivers were produced by entangled pure states. 
    
    The manipulation of quantum information gives access to exponentially large continuous systems (where before we had $n$ bits of classical values, the $n$-bit-strings now index the standard basis of a $2^n$ dimensional vector space) and provably stronger correlations than classical systems (entanglement).
    However, we cannot just access all this information. Everything is hidden behind the no-cloning theorem \cite{QuantumBook2021} and we can only access the information partially via a measurement. In particular still $n$ bits of information can only be collected by measuring $n$ qubits.

\section{The Quantum Control Plane} 
\label{sec:QCNarchitecture}

    Quantum information processing does not give an automatic advantage over classical information processing, and exactly determining for which problems the advantage exists is still an open research.
    Nonetheless, we already have important quantum algorithms of independent use and wide applications, with quantum simulations in chemistry likely becoming the first real applications with an exponential speedup~\cite{preskill2012quantum,Preskill_2018}.
    A quantum network capable of connecting quantum computers will be essential to support these applications of quantum computing and it will rely on the existent classical network infrastructure. 
    Most importantly (the aspect to which we want to bring our attention), quantum computation and communication have the potential to return some benefits to the classical network itself, by reducing overhead, latency, and congestion, while increasing energy efficiency.
        
    Various classes of unsupervised ML and optimisation problems have been proven to have an exponentially lower computational cost, once the expensive step of encoding the large amount of data in the coefficients of a quantum state has been performed~\cite{LMR13,LMR14,RML14,brandao2017sdp}.
    Notice, that these are not exponential speedups in the strict sense, because the gain is obtained by changing the input to the algorithm, thus making the classical and the quantum algorithm incomparable.
    To gain this exponential advantage, the type of encoding is crucial.
    We have seen that we can encode each $n$ bit-string onto a standard basis vector, this encoding is lossless but also uses more expensive $n$-qubit strings.
    Alternatively, we can take any $n$ dimensional vector, normalise it, and construct a quantum state of $\log n$ qubits corresponding to this vector. The encoding is extremely lossy, because by measuring the encoded state we can only obtain $\log n$ bits of information, but also extremely storage efficient.
    For example, a vector of one petabyte with one-byte coefficients can be encoded in just 15 qubits.
    Since in order to construct the quantum state, the classical data must be accessed, this encoding is still an expensive step.
    Once the expensive encoding is done, $k$-means~\cite{LMR13}, principal component analysis (PCA)~\cite{LMR14}, support vector machines (SVM)\cite{RML14}, and any semidefinite program~(SDP) -- a general class of optimisation problems that include linear programs -- can be run efficiently, namely polynomially in the number of qubits, with a quantum computer.
    
    By relying the network optimisation on algorithms that fit this class, we can experience an exponential reduction on the traffic and the computational load introduced by the intelligent control-plane operations.
    Even any other algorithm that relies on the classical data, being encoded in the coefficients of the superposition, which does not have the exponential speed up, will incur an exponential reduction of the traffic load.
    In particular, a quantum convolutional neural network will at least gain access to exponentially deeper networks at the same training cost~\cite{KLP19}.
    The required flexibility of the network and the amount of data analysed make neural network and unsupervised ML algorithms the main choice for ML in the hypervisor~\cite{XieYuHuangXieLiuWangLiu2019}. This makes the quantum algorithms mentioned above particularly relevant for this application.
    
    The expensive quantum encoding of classical Big Data would be distributed and performed on-site at the nodes of the network where the data is collected.
    Namely, each data-source node would collect their data on a few qubits rather than streaming a large number of classical data via the classical network.
    At this point only a logarithmic number of qubits, compared to the number of classical bits that would otherwise be required, needs to be sent to a processing quantum data centre, hosting the hypervisor.
    A large load on the classical network thus becomes a small load on the quantum network.
    
\begin{figure}
	    \centering
		\includegraphics[width=.9\columnwidth]{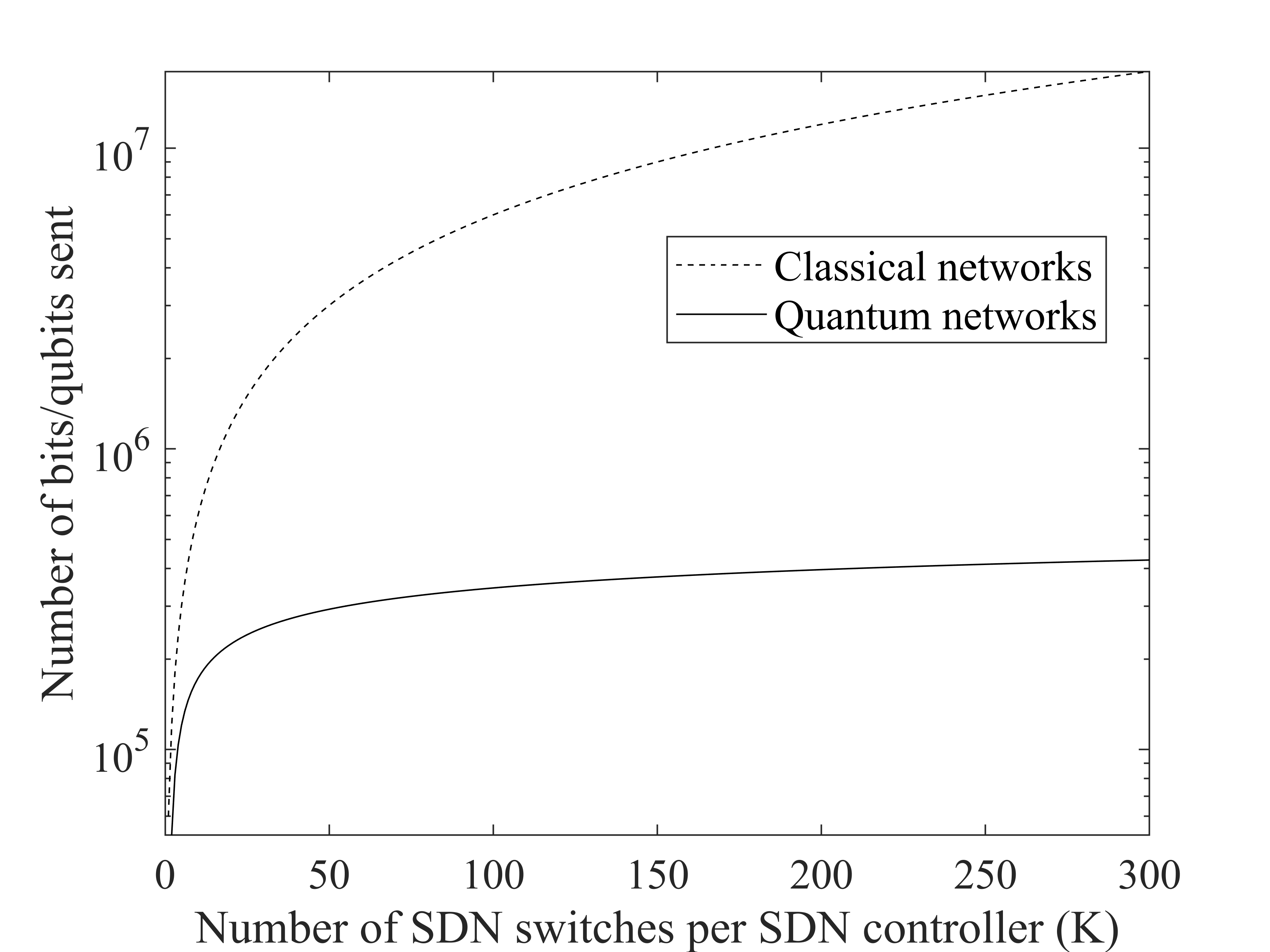}
		\caption{Example of scaling of the number of bits sent to the hypervisor, compared to the number of qubits sent using QRAM. The evaluation assumes each one of the $N$ SDN controllers manages the same number $K$ of SDN switches.}
	    \label{fig:plotlatency}
 	\end{figure}

    To take full advantage of this encoding, the intermediate nodes will have to be capable of using the quantum analog of the random access memory (QRAM)~\cite{GLM08} and all nodes need to be able to encode in superposition a vector of data into a state where each vector coefficient gives a coefficient in the superposition. 
    A typical B5G data-plane node will be controlled by up to $2000 \sim 2^{11}$ parameters~\cite{IZD14} that can be encoded in 11 qubits.
    Via QRAM, the quantum states can be collected at the SDN nodes.
    If the SDN controller receives data from $K$ switches then with $\log K$ qubits the SDN node can use QRAM to join the qubits from all the nodes into $11 + \log K$ qubits.
    Namely, the 11 qubits of all the switches are joined into the same 11 qubits via entanglement with the $\log K$ qubits of QRAM address.
    The nodes then can send the newly encoded quantum states with minimum overhead to the central hypervisor, which will in turn join the states received by the $N$ quantum SDN controllers, into a state of $11 + \log K + \log N$ qubits.
    This advantage is displayed in \cref{fig:plotlatency}.
    The communication at every link is reduced exponentially, compared to each SDN controller sending $K \cdot 2000$ parameters and the central hypervisor receiving $N\cdot K \cdot 2000$ parameters.
    Indeed, if each SDN controller receives $1024$ switches and the SDN controllerhypervisor receives from $1024$ SDN controllers, then the final quantum states only occupies $31$ qubits at the hypervisor.
    The total amount of sent qubits is $11 \cdot 1024 + 21 \cdot 1024$, however distributed over for example $1024 \cdot 1024$ control-plane links, in contrast to the $2000 \cdot 1024 + 2000 \cdot 1024 \cdot 1024$ classical parameters.

    The quantum algorithms, however, are probabilistic and usually work by collecting the expectation value over the repetition of the algorithms many times. 
    Because of the no cloning theorem \cite{QuantumBook2021}, the hypervisor will not be able to generate the copies of the encoded data needed for the repetition, thus the drawback is that for every repetition the data must be re-encoded in the quantum states at the source.
    If for example $100$ samples are needed, then the whole operation must be repeated $100$ times, thus multiplying by $100$ the amount of sent qubits.
    However, the number of repetition is generally the inverse polynomial on an approximation/error parameter and thus as the number of parameters grows exponentially with future generation networks, the number of repetitions will scale polynomially.
    Namely, the quantum algorithm is scalable to the increasing size and complexity of future generation networks.
    Already at the B5G nodes with $2000$ parameters, this quantum encoding allows $200$ samples to be taken before the quantum communication exceeds the needed classical communication \emph{at the leaf nodes}. The bandwidth reduction within the control-plane links is exponentially larger if we consider nodes that send all data to the hypervisor. 
    If instead we consider SDN controllers that classically reduce the data via some ML preprocessing before sending it to the hypervisor, the bandwidth reduction might be comparable.
    However, in this comparison the quantum ML algorithm at the hypervisor has a full-global view of the network status, while the classical algorithm has a reduced and indirect view of the global status, depending on the reduced data.

    This reduction of packets sent, stored and processed within the control plane will reduce congestion and reduce almost all sources of delay such as queuing and transmission delay in the network, and storage and processing delay at the processing centre.
    As we have seen, each additional load to one layer also contributes to load the other.
    In particular, each additional processing and softwarization itself introduces new delays and loads on other layers of the network. 
    Thus the smaller queues due to the quantum processing further contributes to the reduction of processing and communication needed for queues' management.
    The processing required by the ML algorithm at the hypervisor to handle Big Data of the network will consume significant energy. In general, the amount of energy saving directly depends on the number of bytes sent to the hypervisor and the number of instructions for computations, and inversely depends on the available bandwidth at the communication links~\cite{CaoLiuChengShen2018}. Thus, the reduction due to quantum deployment described above will not only reduce latency but it will also imply higher energy efficiency (in line with 5G and B5G goals).

    Furthermore, shared entanglement will allow the classical and the quantum networks to transfer each others' load.
    Namely, stored entanglement allows to exchange quantum and classical communication through the teleportation and dense-coding protocols, in order to balance the load in the hybrid network. 
    In quantum teleportation, one qubit can be communicated using previously-shared entanglement and sending two (classical) bits.
    In dense-coding two bits of communication can be achieved by sending one qubit of the previously-shared entanglement.
    This is one additional reason to consider obsolete the static-layer view of the network.
    The classical and quantum networks cannot be two distinct layers of the network that are optimised independently because in this way the flexibility we just described is lost.
    
    Finally, shared entanglement has the potential of even reducing communication complexity, by allowing nodes in the network to send data that is more correlated than classically to the processing centres.
    This is not the case of extracting a small amount of information from large data, and is not in contradiction with a measurement on $n$ qubits only being able to provide $n$ bits of information.
    Sometimes the required amount of communication to perform a task is less than the communication needed to solve the same task classically, as demonstrated by non-local games such as the Clauser-Horne-Shimony-Holt (CHSH) and the Mermin game~\cite{QuantumBook2021}.
    However, for this advantage to exist, strong non-local correlations between the environments, adversarial setting, or strong limitations on the type of channels must be present.
    Therefore further research is needed to determine whether such advantages are realistic.

\section{Conclusion}

    We have seen that quantum computation and communication will not necessarily be only a new resource but can also contribute to reduce overheads, latency, and increase energy efficiency in the classical \emph{softwarized} networks. However, quantum information processing does not \emph{magically} improve any computation or communication tasks and only a few classes of problems have been shown to take advantage from quantum technologies. Notably, Grover's search algorithm \cite{QuantumBook2021} is an extremely general algorithm, capable of providing up to quadratic quantum speedup on almost any algorithm, that will surely find its applications in every field, including future \emph{softwarized} networks. Our aim was to clearly point out where there is a huge potential for future networks to benefit from quantum technologies.
    In parallel, even the classical part of future network infrastructures should be designed with quantum technologies in mind, to efficiently and effectively merge the two into a unique hybrid architecture.
    The use of quantum technologies in computation and communication is a young field but is now quickly maturing so that possible applications will appear in the near future\footnote{See the European Quantum Flagship \url{https://qt.eu/} (Feb. 2021).}.
    Thus, the final goal is to see a paradigm shift in the network development, where classical and quantum technologies become two sides of the same hybrid communication network.


\section*{Acknowledgements}
    
    This work has been partially funded by the German Research Foundation (DFG, Deutsche Forschungsgemeinschaft) as part of Germany’s Excellence Strategy – EXC2050/1 – Project ID 390696704 – Cluster of Excellence “Centre for Tactile Internet with Human-in-the-Loop” (CeTI) of Technische Universität Dresden.
    Holger Boche, Christian Deppe and Roberto Ferrara were supported by the Bundesministerium f\"ur Bildung und Forschung (BMBF) through Grants 16KIS0856 (Deppe, Ferrara), 16KIS0858 (Boche), and 16KIS0948 (Boche). This work has also been partially funded by the Deutsche Forschungsgemeinschaft (DFG, German Research Foundation) within Germany’s Excellence Strategy under Grant EXC-2111 390814868.


\vspace{-4em}
\begin{IEEEbiographynophoto}{Roberto Ferrara}
obtained his Ph.D. in science at the Department of Mathematical Sciences of the University of Copenhagen. Since 2019 he is at the Department of Communications Engineering at the Technische Universität München.
\end{IEEEbiographynophoto}
%
\begin{IEEEbiographynophoto}{Riccardo Bassoli}
is  a  senior  researcher  with  the  Deutsche  Telekom  Chair  of  Communication  Networks  at  the  Faculty  of  Electrical and Computer Engineering, Technische Universität Dresden (Germany). He received his Ph.D. degree from 5G Innovation Centre (5GIC) at University of Surrey (UK), in 2016.
\end{IEEEbiographynophoto}
%
\begin{IEEEbiographynophoto}{Christian Deppe}
received the Dr.-Math. degree in mathematics from the Universität
Bielefeld, Bielefeld in 1998. He was a Research and Teaching Assistant with the
Universität Bielefeld. Since 2018 he is at the Department of Communications
Engineering, Technical University of Munich, Munich, Germany.
\end{IEEEbiographynophoto}
%
\begin{IEEEbiographynophoto}{Frank H.P. Fitzek}
received  the Dr.-Ing. degree in Electrical Engineering  from  the Technical University Berlin, Germany. He is  a  Professor  and  head  of the  Deutsche Telekom Chair of Communication Networks at Technische Universität Dresden.
\end{IEEEbiographynophoto}
%
\begin{IEEEbiographynophoto}{Holger Boche}
received the Dr.-Ing. degree in electrical engineering from the Technische Universität Dresden in 1990 and 1994, respectively, and the Dr. rer. nat. degree in pure mathematics from the Technische Universität Berlin in 1998. Since 2010, he is Full Professor at the Technische Universität München, Munich, Germany.
\end{IEEEbiographynophoto}

\end{document}